\documentclass[onecolumn,
floatfix,%
reprint,
superscriptaddress,
 amsmath,amssymb,
 aps,
]{revtex4-2}

\usepackage{multirow}
\usepackage{siunitx}
\usepackage{graphicx}% Include figure files
\usepackage{dcolumn}% Align table columns on decimal point
\usepackage{bm}% bold math
\begin{document}

%\preprint{APS/123-QED}

\title{Neutron capture measurement of the $^{165}$Ho at the CSNS Back$-$n facility in the resonance energy region}% Force line breaks with \\

\author{De-Xin Wang}
\affiliation{College of Physics and Electronics, and Institute of Nuclear Physics, Inner Mongolia Minzu University, Tongliao $028043$, China}
\affiliation{Inner Mongolia Joint Key Laboratory of Nuclear and Radiation Detection,Tongliao $028043$, China}

\author{Su-Ya-La-Tu Zhang}
\email[Corresponding author, ]{ E-mail addresses: zsylt0416@163.com}
\affiliation{College of Physics and Electronics,  and Institute of Nuclear Physics, Inner Mongolia Minzu University, Tongliao $028043$, China}
\affiliation{Inner Mongolia Joint Key Laboratory of Nuclear and Radiation Detection,Tongliao $028043$, China}

\author{Wei Jiang}
\affiliation{Institute of High Energy Physics, Chinese Academy of Sciences, Beijing $10049$, China }
\affiliation{Spallation Neutron Source Science Center, Dongguan $523803$, China }

\author{Rui-Rui Fan}
\affiliation{Institute of High Energy Physics, Chinese Academy of Sciences, Beijing $10049$, China }
\affiliation{Spallation Neutron Source Science Center, Dongguan $523803$, China }

\author{Qi-Wei Zhang}
\email[Corresponding author, ]{ E-mail addresses: zqwvictor@126.com}
\affiliation{Key Laboratory of Nuclear Data, China Institute of Atomic Energy, Beijing $102413$, China }

\author{Jie Ren}
\affiliation{Key Laboratory of Nuclear Data, China Institute of Atomic Energy, Beijing $102413$, China }

\author{Jin-Cheng Wang}
\affiliation{College of Science, Inner Mongolia University of Technology, Hohhot $010051$, China}

\author{Guang-Yuan Luan}
\affiliation{Key Laboratory of Nuclear Data, China Institute of Atomic Energy, Beijing $102413$, China }

\author{Xiao-Guang Wu}
\affiliation{Key Laboratory of Nuclear Data, China Institute of Atomic Energy, Beijing $102413$, China }

\author{Bao-Hua Sun}
\affiliation{ School of Physics, Beihang University, Beijing $10019$, China }

\author{Zhen-Xiang Zhou}
\affiliation{School of Physics and State Key Laboratory of Nuclear Physics and Technology, Peking University, Beijing $100871$, China}

\author{Hong-Yi Wu}
\affiliation{School of Physics and State Key Laboratory of Nuclear Physics and Technology, Peking University, Beijing $100871$, China}

\author{Zhi-Yang He}
\affiliation{Key Laboratory of Nuclear Data, China Institute of Atomic Energy, Beijing $102413$, China }

\author{Cong-Bo Li}
\affiliation{Key Laboratory of Nuclear Data, China Institute of Atomic Energy, Beijing $102413$, China }

\author{Qi Sun}
\affiliation{Key Laboratory of Nuclear Data, China Institute of Atomic Energy, Beijing $102413$, China }

\author{Xuan Pang}
\affiliation{ School of Physics, Beihang University, Beijing $10019$, China }

\author{Mei-Rong Huang}
\affiliation{College of Physics and Electronics,  and Institute of Nuclear Physics, Inner Mongolia Minzu University, Tongliao $028043$, China}
\affiliation{Inner Mongolia Joint Key Laboratory of Nuclear and Radiation Detection,Tongliao $028043$, China}

\author{Guo Li}
\affiliation{College of Physics and Electronics,  and Institute of Nuclear Physics, Inner Mongolia Minzu University, Tongliao $028043$, China}
\affiliation{Inner Mongolia Joint Key Laboratory of Nuclear and Radiation Detection,Tongliao $028043$, China}

\author{Gerile Bao}
\affiliation{School of Control and Computer Engineering, North China Electric Power University, Baoding $071003$, China}

\author{Xi-Chao Ruan}
\affiliation{Key Laboratory of Nuclear Data, China Institute of Atomic Energy, Beijing $102413$, China }

\begin{abstract}

The neutron capture yield of $^{165}$Ho was measured at the Back-streaming White neutron beam line (Back-n) of the China Spallation Neutron Source (CSNS) using a 4$\pi$ BaF$_2$ Gamma Total Absorption Facility (GTAF). The resonance shapes in the energy range from \SI{1}{\electronvolt} to \SI{1.0}{k\electronvolt} were analyzed using the Bayesian R-matrix code SAMMY. For 18 s-wave resonances below \SI{100}{\electronvolt}, the resonance energy $E_R$, neutron width $\Gamma_n$, and radiative width $\Gamma_{\gamma}$ were extracted. The statistical analyses of the resonance parameters show that the nearest-neighbor level-spacing distribution follows a Wigner–Dyson form with mean spacing $D_0 = 4.53(3)$ eV, indicating chaotic compound-nucleus behavior. The reduced neutron widths follow the Porter–Thomas $\chi^{2}$ distribution with one degree of freedom ($v = 1$), consistent with a single entrance channel. 
The mean radiative width for s-wave resonances was determined to be $\langle\Gamma_{\gamma}\rangle$ = $88.10 \pm 1.98$ $\text{meV}$ using the weighted-average technique.
From the extracted parameters, the s-wave neutron strength function for $^{165}$Ho was derived to be $10^{-4}S_0=\SI{2.01 \pm 0.01}{}$, in excellent agreement with values reported in both the Atlas of Neutron Resonances and ENDF/B-VIII.0 data.

\begin{description}
\item[PACS numbers]
25.70.Pq. 
\item[Keywords]
Neutron Capture; Total Absorption Calorimeter; CSNS; $^{165}$Ho(n,$\gamma$).

\end{description}
\end{abstract}

\maketitle

%\tableofcontents

\section{\label{sec:level1}Introduction}
Neutron capture data are indispensable for modern nuclear data evaluations because they constrain reaction models and underpin a wide range of applications, from reactor design to astrophysical nucleosynthesis\cite{1959MossinKotin}. For decades, the 
$^{197}$Au(n,$\gamma$) reaction has served as the primary standard, thanks to carefully validated cross-section measurements performed at many laboratories \cite{1988Ratynski,2009Carlson,2010Massimi,2011Lederer}. Nevertheless, several studies have suggested that $^{165}$Ho(n,$\gamma$) possesses properties that could make it a complementary standard in specific experimental configurations \cite{1973czirr}.
Holmium shares several favorable features with gold: it exhibits a well resolved and representative prompt-$\gamma$ spectrum following neutron capture \cite{1965Brunhart}, its product nucleus has a convenient half-life for offline studies, and, most importantly, it features a pronounced s-wave “black” resonance at \SI{3.9}{\electronvolt} with an exceptionally small scattering ratio (\(\Gamma_{n}/\Gamma_T = 0.025\))\cite{1976Macklin}. Because virtually all incident neutrons are absorbed at this energy, corrections for multiple scattering are minimal, greatly simplifying the absolute normalization of capture yields \cite{1982Alfimenkov}. Despite these advantages, the adoption of $^{165}$Ho(n,$\gamma$) as a secondary standard has been impeded by the scarcity of high precision capture data below in the resonance region.

\begin{figure}[!htb]
\includegraphics[width=.95\linewidth]{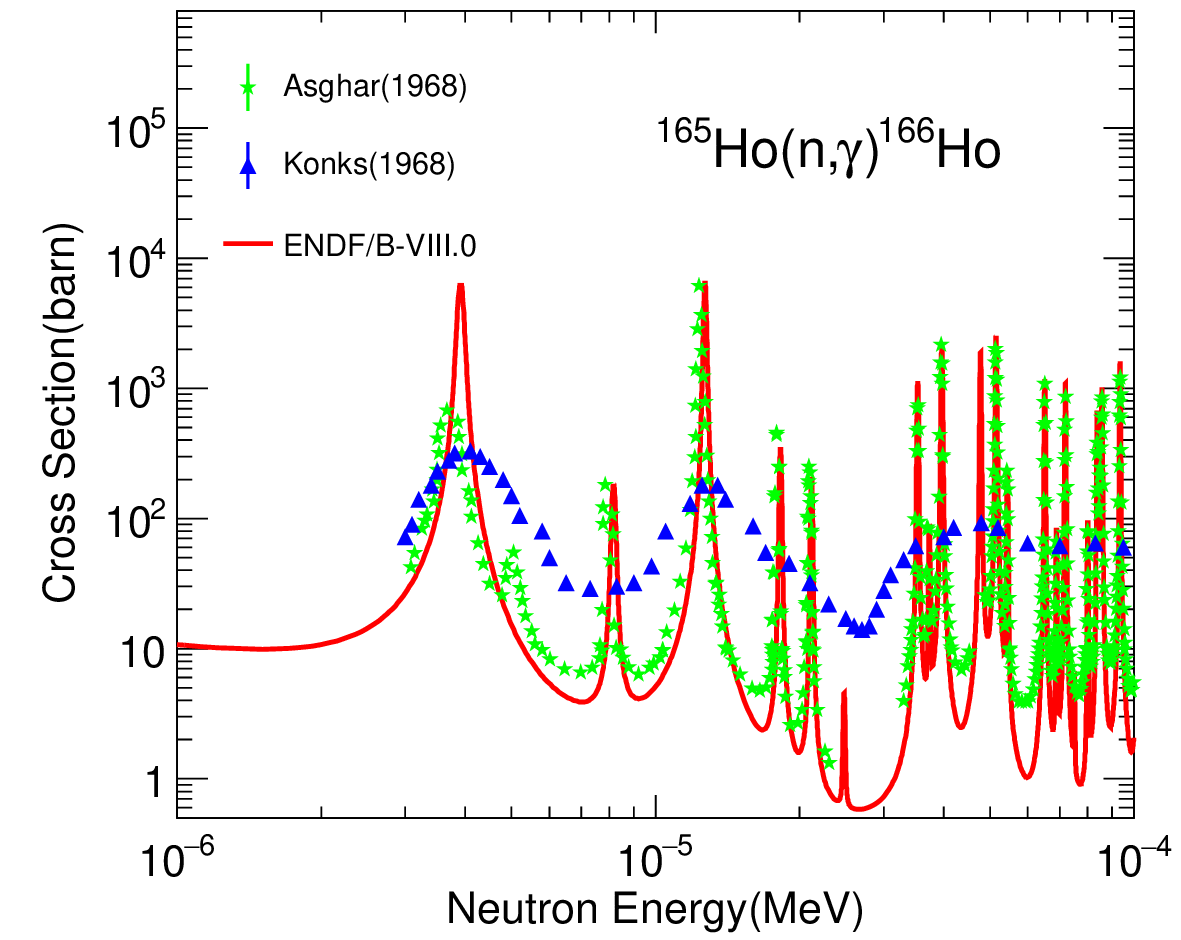}
\caption{(Color online) In the resonance energy region of $^{165}$Ho neutron capture cross section of ENDF/B-0VIII.0 compared with the data from Asghar {\it et al.}~\cite{1968Asghar} and Konks {\it et al.}~\cite{1968Konk}. }
\label{fig:cross section}
\end{figure}

From the standpoint of nuclear structure physics, new data on $^{165}$Ho(n,$\gamma$) are equally compelling. Odd-odd nuclei such as $^{166}$Ho remain a stringent test bed for microscopic approaches shell model calculations, collective models, and the interacting boson framework still face difficulties in reproducing their level schemes and decay patterns\cite{2003Bender}. Radiative capture experiments provide resonance energies, neutron widths, and radiative widths that are sensitive to the single particle and collective components of the wavefunction\cite{1967Motz,2023Pogliano}. At present, reliable experimental information for $^{165}$Ho(n,$\gamma$) in the resolved resonance region is very limited, and marked discrepancies persist between evaluations and the few existing measurements~\cite{1968Asghar,1968Konk}, as illustrated in Fig.~\ref{fig:cross section}.

%\vspace{-0.4cm}
%\vspace{-0.4cm}

Recent developments in time-of-flight facilities, high-granularity $\gamma$  detectors, and digital data acquisition systems now make it possible to revisit $^{165}$Ho(n,$\gamma$) with substantially improved resolution and statistical precision. The present work therefore reports new neutron capture measurements on $^{165}$Ho with the Gamma Total Absorption Facility (GTAF)~\cite{2021Zhang}, a 4$\pi$ $\gamma$-ray detector array consisting of 40 BaF$_2$ modules operating at the Back-streaming White Neutron beam line (Back-n) at the China Spallation Neutron Source (CSNS)~\cite{2022Luan}. Section II details the experimental setup, samples, and associated parameters. Section III outlines the methodology for extracting capture yields from the experimental data. In Section IV, we present the results of the capture yields, including a resonance shape analysis with the R-matrix code SAMMY~\cite{OakRidgeReport}, and statistical properties of the average resonance parameters. Finally, Section V summarizes the conclusions of the present work and its implications for future studies.

\section{\label{sec:level2}Measurement}
%\subsection{The Back-n facility at CSNS}
The Back-n facility at the CSNS is a time-of-flight (TOF) facility designed for high-precision neutron-induced reaction measurements~\cite{2021Tang,2021LiXX}. The Back-n facility is located in the opposite direction of the CSNS proton beam and extends about \SI{80}{\metre} from the spallation target, providing a broad neutron beam ranging from \SI{0.3}{\electronvolt} to \SI{200}{\mega\electronvolt}~\cite{2021CPCTang,2021Hu}. After several upgrades, the operating power has now reached 170 kW. 
The unique countercurrent geometry of the Back-n minimizes the background contribution of forward-scattered particles, enhancing its applicability to low-energy neutron reaction measurements, which are critical for nuclear data studies~\cite{2022wang,2023ZhangS}.

The Back-n beamline has two experimental stations, ES\#1 and ES\#2, located at flight path lengths of \SI{55}{\metre} and \SI{76}{\metre}, respectively. ES\#2, the station for present work, is used for neutron capture experiments and offers a neutron flux of $6.92\times10^{5}\text{cm}^{-2} \cdot \text{s}^{-1}$ at the sample position~\cite{2021Tang}. This flux is distributed over three configurable beam spot sizes— $\Phi$\SI{20}{\mm}, $\Phi$\SI{30}{\mm}, and $\Phi$\SI{60}{\mm}~\cite{2019chen}. The neutron energy spectrum at ES\#2 was characterized using two detection systems: a lithium-silicon (Li-Si) detector based on the $^{6}$Li(n,t)$^{4}$He reaction and a $^{235}$U fission chamber exploiting the $^{235}$U(n,f) reaction~\cite{2023chen}. 
A silicon flux monitor (SiMon) located upstream of the sample \SI{20}{\metre} continuously monitors beam intensity stability. The SiMon consists of a thin $^{6}$LiF conversion layer coupled with eight silicon detectors arranged outside the beam path, ensuring real-time flux normalization without perturbing the neutron beam~\cite{2025chen}.

\begin{figure}[!htb]
\includegraphics[width=.95\linewidth]{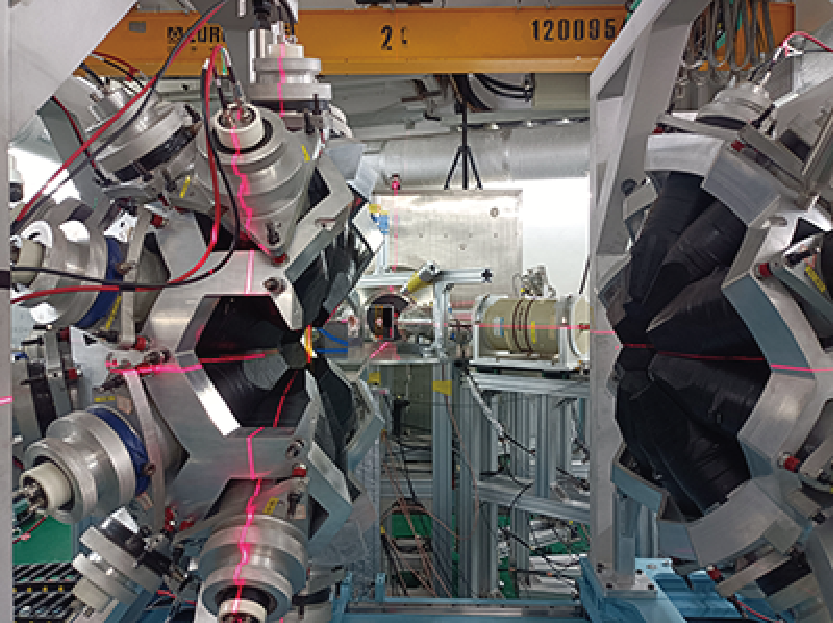}
\caption{(Color online) The setup of the GTAF in the measurement.}
\label{fig:experiment}
\end{figure}

The GTAF is a $4\pi$ array of 40 BaF$_2$ detectors located at ES\#2 (as shown in Fig.~\ref{fig:experiment}), designed for total‑absorption $\gamma$-ray spectroscopy\cite{2021Zhang}. The array forms a spherical shell comprising 12 pentagonal and 28 hexagonal frustum‑shaped BaF$_2$ crystals, with an inner diameter of 20 cm and a radial thickness of 15 cm. The energy calibration of each $\mathrm{BaF}_2$ detector was conducted using standard $\gamma$-ray sources, including $^{60}$Co (\SI{1.173}{M\electronvolt} and \SI{1.332}{M\electronvolt}) and $^{137}$Cs (\SI{0.662}{M\electronvolt}) in the experiment. The signals from each detectors are collected with the XIA digital data acquisition system (DAQ) developed by the Experimental Nuclear Physics Group of Peking University~\cite{2020Wu}. To prevent the massive amount of experimental data generated by GTAF, the XIA DAQ does not record the waveforms in real-time, but rather extracts important information such as the detector number, energy spectrum, and trigger time from the output waveform. The data analysis was conducted using the ROOT software, developed by the European Organization for Nuclear Research (CERN)~\cite{root}.

%\vspace{-0.3cm}
%\subsection{The $^{165}$Ho and samples}
A metallic sample of holmium ($^{165}$Ho) with a diameter of \SI{30.0}{mm} and an area density of $6.42\times10^{-4}$ atoms/barn was utilized for the measurement of prompt capture $\gamma$ rays. The sample, with a purity of 99.5\% (containing 0.5\% $^{nat}$Dy), was prepared at the China Institute of Atomic Energy. 
The background of the empty target was measured, while the effects of scattered neutrons were analyzed using a sample of $^{nat}$C.
The details of the sample used are listed in Table~\ref{tab:Sample}. In this work, the measurement durations for different samples were set as follows: the $^{165}$Ho sample was measured for 11 hours to obtain sufficient capture event statistics. The $^{nat}$C target, used for scattering background analysis, was measured for 3 hours. The empty sample holder was measured for 12 hours to accurately characterize the inherent background. 

\vspace{-0.6cm}
\begin{table}[hb]%The best place to locate the table environment is directly after its first reference in text
\caption{
\label{tab:Sample}%
Sample information
}
\begin{ruledtabular}
\begin{tabular}{lcdr}
\multirow{2}{*}{sample} &Thickness&Mass&Area density  \\
  & /mm& /mg& /$atom\cdot b^{-1}$ \\
\colrule
$^{165}$Ho  & 0.2      & 1243.32  &$6.420\times10^{-4}$     \\
$^{nat}$C  & 1.0      & 1597.5 &$1.134\times10^{-2}$  \\
Empty holder  &  &      &      \\
\end{tabular}
\end{ruledtabular}
\end{table}

\vspace{-0.5cm}
\section{Data analysis}
The neutron kinetic energy is calculated from the time of flight using Eq.(\ref{eq1}):
\begin{small}
\begin{eqnarray}
E_n= \frac{1}{2} m_n (\frac{L_0 + \Delta L(E_n)}{t-t_0})^2
\label{eq1}
\end{eqnarray}
\end{small}
where $m_n$ is the neutron mass, $L_0$ is the neutron flight distance from the spallation target to the detector, $\Delta L(E_n)$ is the moderation length, t is the detector's observed time of neutrons, and $t_0$ is the observed time of $\gamma$-flash. 
The capture yield, derived from neutron energy and experimental data, is calculated using Eq.(\ref{eq2})~\cite{2021Zhang}.
\begin{small}
\begin{eqnarray}
Y_{exp}(E_n)= \frac{C(E_n)-B(E_n)}{N \varepsilon \Phi (E_n)}
\label{eq2}
\end{eqnarray}
\end{small}
where $E_n$ is the neutron energy, $C(E_n)$ and $B(E_n)$ represent the net counts of the sample and background, respectively. N is the normalization factor for the capture yield, and $\varepsilon$ is the detector efficiency, $\Phi (E_n)$ is the neutron flux. In this work, the experimental data were absolutely normalized to $N$ using the saturated resonance peak $^{165}$Ho at 12.7 eV.

When the neutron capture cross section ($\sigma_c$) and the total neutron cross section ($\sigma_t$) are known, the capture yield can be further refined using Eq.(\ref{eq3}):
\begin{small}
\begin{eqnarray}
Y_{th}(E_n)=(1-e^{n \sigma_t(E_n)}) \frac{\sigma_c(E_n)}{\sigma_t (E_n)}
\label{eq3}
\end{eqnarray}
\end{small}
where $n$ is the sample atom density.

\subsection{Background subtractions}
\begin{figure}[!htb]
\includegraphics[width=.95\linewidth]{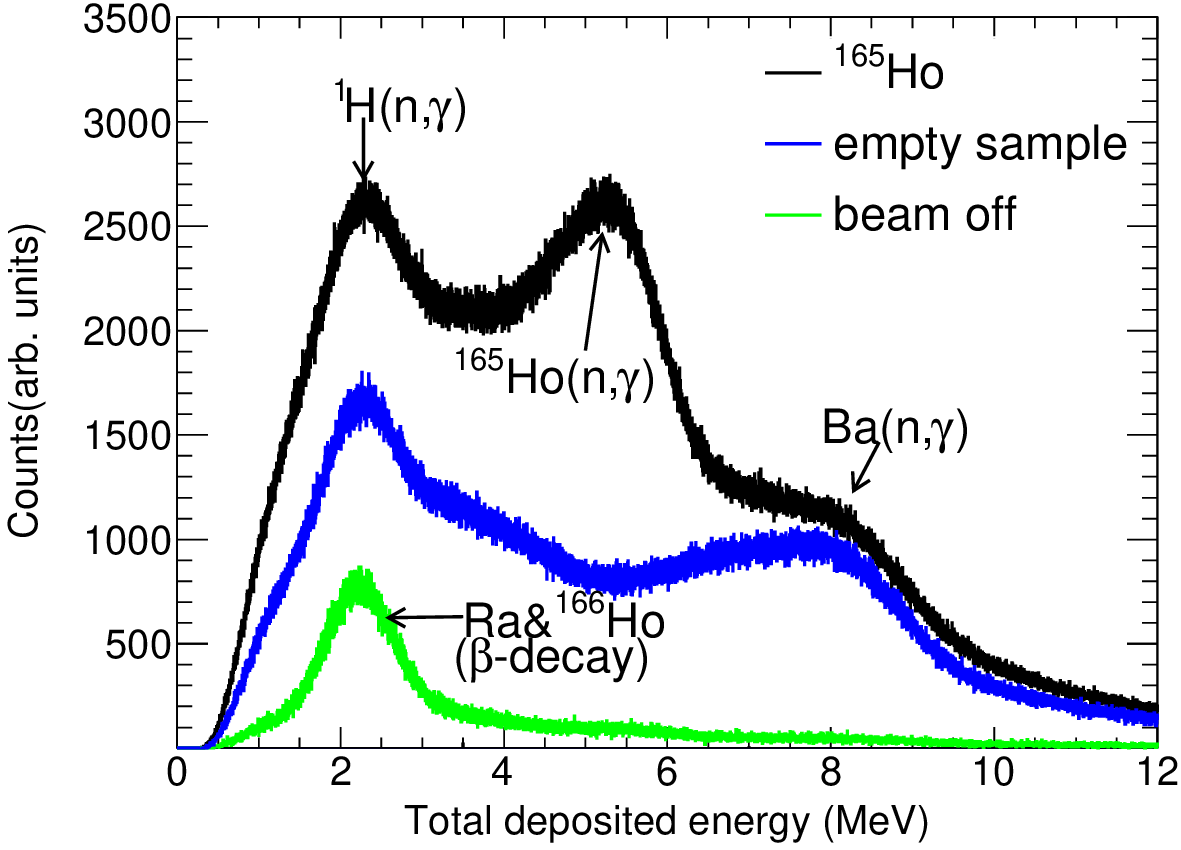}
\caption{(Color online) Total deposited energy $E_{sum}$ for $^{165}$Ho, empty target and beam off.}
\label{fig:Counting}
\end{figure}

The background in the capture measurement was determined under three conditions: a $^{165}$Ho sample, an empty target, and beam-off. The response of the GTAF to these samples is shown in Fig.\ref{fig:Counting}. 
The detector response to the $^{165}$Ho capture cascade reaction exhibits a distinct peak at 5 MeV, corresponding to the total absorption of the cascade $\gamma$ rays.

From Fig.\ref{fig:Counting}, it can be observed that the components of the background mainly include: 2.2 MeV $\gamma$ rays generated by the moderator water around the spallation target through the $^{1}$H(n,$\gamma$)$^{2}$H reaction~\cite{2018Jing}, $\gamma$ rays between 7 MeV and 9 MeV generated by the Ba (n, $\gamma $) reaction inside the BaF$_2$ detector, and low-energy background $\gamma$ rays originating from radioactive samples ($^{166}$Ho), activated products ($^{27}$Al,$^{208}$Ti), radium impurities in the BaF$_2$ detector, and the environment background ($^{40}$K)~\cite{2009Guerrero}.
They are categorized according to their correlation with the neutron beam, leading to two different types: in-beam and beam-off.
The reaction types and Q values of the background are detailed in Table \ref{tab:Qvalues}.

\begin{table}[!htb]
\caption{Reaction types and corresponding Q values in the background}
\label{tab:Qvalues}
\begin{ruledtabular}
\begin{tabular*}{8cm} {@{\extracolsep{\fill} } lll}
Background &Reaction&Q-value(MeV)\\
\hline
\multirow{3}*{Beam-off} &$^{40}$K($\beta$-decay)    & 1.46  \\
     &$^{208}$Tl($\beta$-decay)   & 2.6 \\
     &$^{226}$Ra(decay-chain)   & 4.8-8.8($\alpha$)\\
\hline
\multirow{3}*{In-Beam} &$^{1}$H(n,$\gamma$)    & 2.2  \\
                       &$^{nat}$Ba(n, $\gamma$)   & 4.7-9.1 \\
                       &$^{27}$Al(n,$\gamma$)   & 7.7\\
\end{tabular*}
\end{ruledtabular}
\end{table}

To reduce the impact of background, a $^{10}$B neutron absorber is usually used between the target and the detector.
Facilities such as TAC~\cite{2018Mendoza} and DANCE~\cite{2005Seyfarth} make use of $^{10}$B spheres to efficiently capture scattered neutrons using the $^{10}$B(n,$\alpha$)$^{7}$Li reaction. 
However, in this work, experimental constraints precluded the use of a neutron absorber. Therefore, optimized crystal multiplicity $m_{cr}$ and total deposited energy $E_{sum}$ thresholds were used to reduce the in-beam background effects.

\begin{figure}[!htb]
\includegraphics[width=1.05\linewidth]{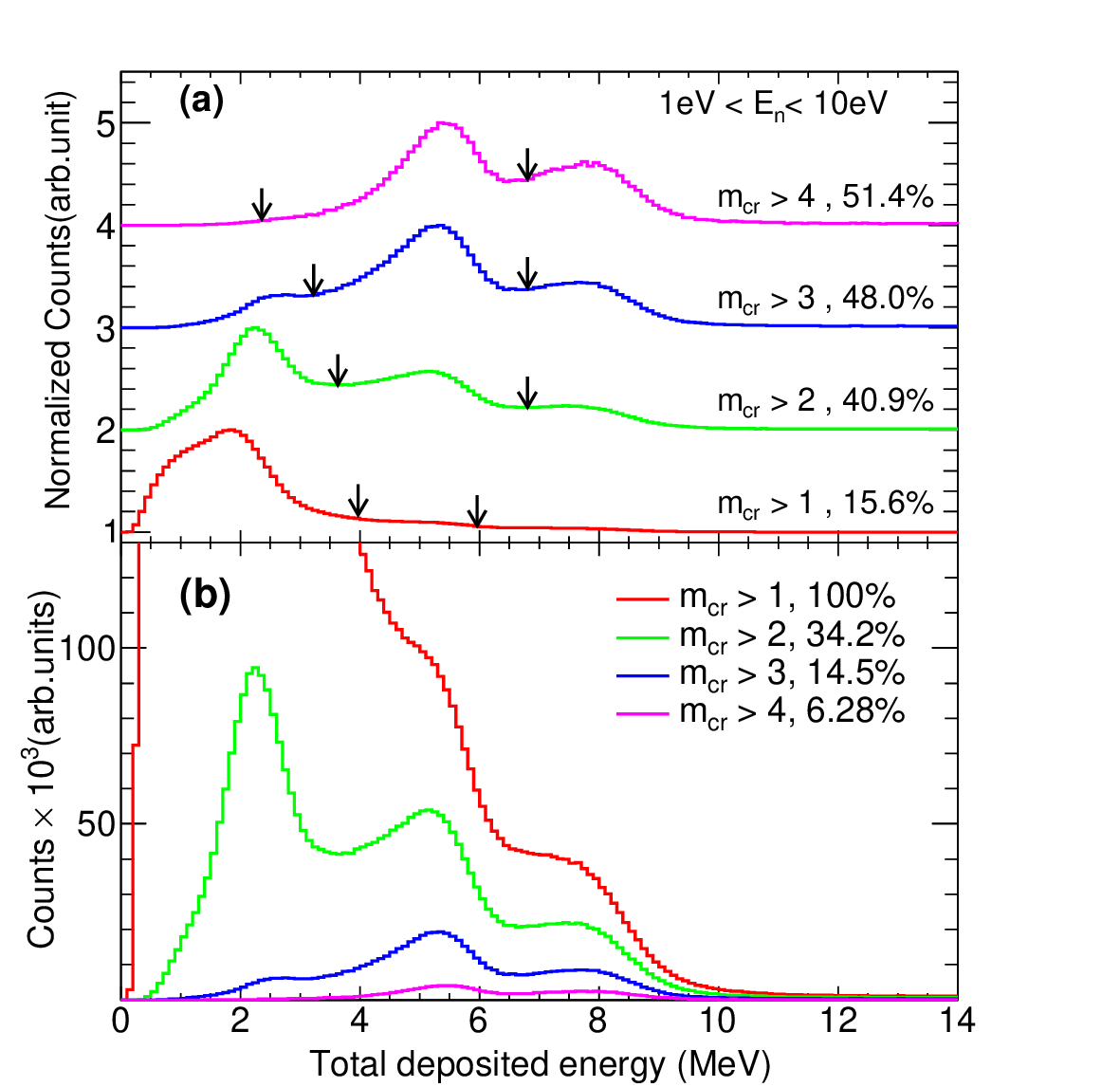}
\caption{(Color online) (a)The $E_{sum}$ corresponding to various $m_{cr}$ ranges with normalization performed relative to the maximum value in the $1eV \textless E_n \textless 10eV$, the area within the two arrows divided by the total area was recognized as the capture-to-background ratio. (b) The $E_{sum}$ distribution of the $^{165}$Ho(n,$\gamma$) measurement in various $m_{cr}$ ranges.}
\label{fig:corresponding}
\end{figure}
For the GTAF event reconstruction process, a 50 ns time overlap window was used to correlate overlapping events. This information is then processed for subsequent data analysis, yielding three key observables:(1)The neutron energy $E_n$, determined from the neutron flight time by the time-of-flight method; (2)Crystal multiplicity $m_{cr}$, representing the number of detected crystal weights for each cascade event; (3)Total deposited energy $E_{sum}$, corresponding to the total energy deposited across all detectors in each cascade.

\begin{figure}[!htb]
\includegraphics[width=.95\linewidth]{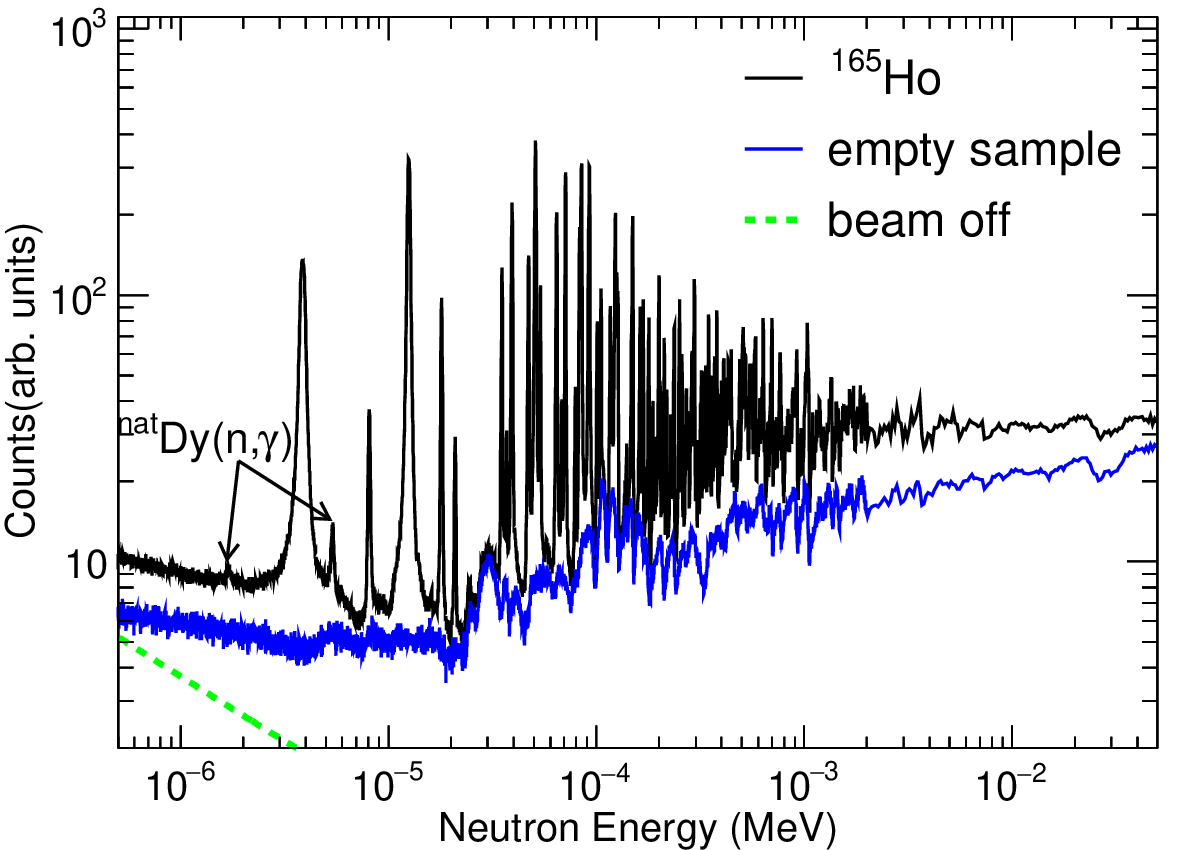}
\caption{(Color online) Comparisons between the $^{165}$Ho spectrum, empty sample and beam off spectrum.}
\label{fig:CuttingCondition}
\end{figure}

A key advantage of the GTAF system is its ability to effectively distinguish and subtract background by using the $m_{cr}$ and $E_{sum}$ thresholds. The normalized $E_{sum}$ for various $m_{cr}$ ranges in the $1eV \textless E_n \textless 10eV$, with normalization performed relative to the maximum value of each spectrum, as shown in Fig.\ref{fig:corresponding}(a).
It can be observed that for $m_{cr}\textgreater 1$, the low-energy background is significant, resulting in a capture-to-background ratio (CBR) of only 15.6\%, as the maximum area corresponds to capture events. This obscures the $^{165}$Ho(n,$\gamma$) peak, typically located in the 4 MeV-6 MeV. 
At $m_{cr}\textgreater 2$, the majority of the low-energy background is suppressed, though a 2.2 MeV $\gamma$-ray peak from neutron capture in the beam becomes more prominent, improving the CBR to 40.9\%.
For $m_{cr}\textgreater 3$, the low-energy background is nearly eliminated, and the $^{165}$Ho(n,$\gamma$) capture peak dominates, achieving a CBR of 48.0\%; however, the event statistics drop by an order of magnitude compared to $m_{cr} \textgreater 1$. 
At $m_{cr} \textgreater 4$, while the $^{165}$Ho(n,$\gamma$) peak remains prominent, contributions from the Ba(n,$\gamma$) reaction become notable.
The total deposited energy of $^{165}$Ho(n,$\gamma$) as a function of $m_{cr}$ is shown in Fig.\ref{fig:corresponding}(b). Setting the relative area for $m_{cr} \textgreater 1$ to 100\%, the areas for $m_{cr}\textgreater 2, 3, 4$ are 34.2\%, 14.5\%, and 6.28\%, respectively.
By comparing Fig.\ref{fig:corresponding}(a) and \ref{fig:corresponding}(b), it can be seen that the $m_{cr} \textgreater 2$ effectively eliminates most background noise while retaining substantial capture data.
Therefore, the condition of $m_{cr} \textgreater 2$ and $3.5 \text{MeV} \textless E_{sum} \textless 7 \text{MeV}$ was used to the present data reduction to minimize background and improve the CBR. Figure \ref{fig:CuttingCondition} shows the $^{165}$Ho spectrum, compared to backgrounds due to the empty sample holder and due to ambient activity. The detection efficiency $\epsilon$ of the GTAF also varies depending on the $m_{cr}$ and $E_{sum}$ conditions. G. Luan et al.~\cite{2021Luan} determined the $\epsilon$ = 89.6 $\pm$ 0.4\% in the $0.05\text{MeV} \textless E_{sum} \textless 9\text{MeV}$ with the mono-energetic $\gamma$ source measurement and Monte Carlo simulations. In this work, the $\epsilon$ is estimated to be about 14.3$\pm$0.06\% for the condition of $m_{cr} \textgreater 2$ and $3.5 \text{MeV} \textless E_{sum} \textless 7 \text{MeV}$.

\vspace{-0.5cm}

\subsection{Neutron sensitivity}
In neutron capture experiments, maintaining low neutron sensitivity is essential. 
Neutron sensitivity is defined as the efficiency of detecting scattered neutrons.
This reduces background contributions from scattered neutrons and prevents systematic errors in resonance parameter determination, particularly when the capture reaction structurally resembles resonance scattering~\cite{2012Guerrero}. 

\begin{figure}[!htb]
\includegraphics[width=.85\linewidth]{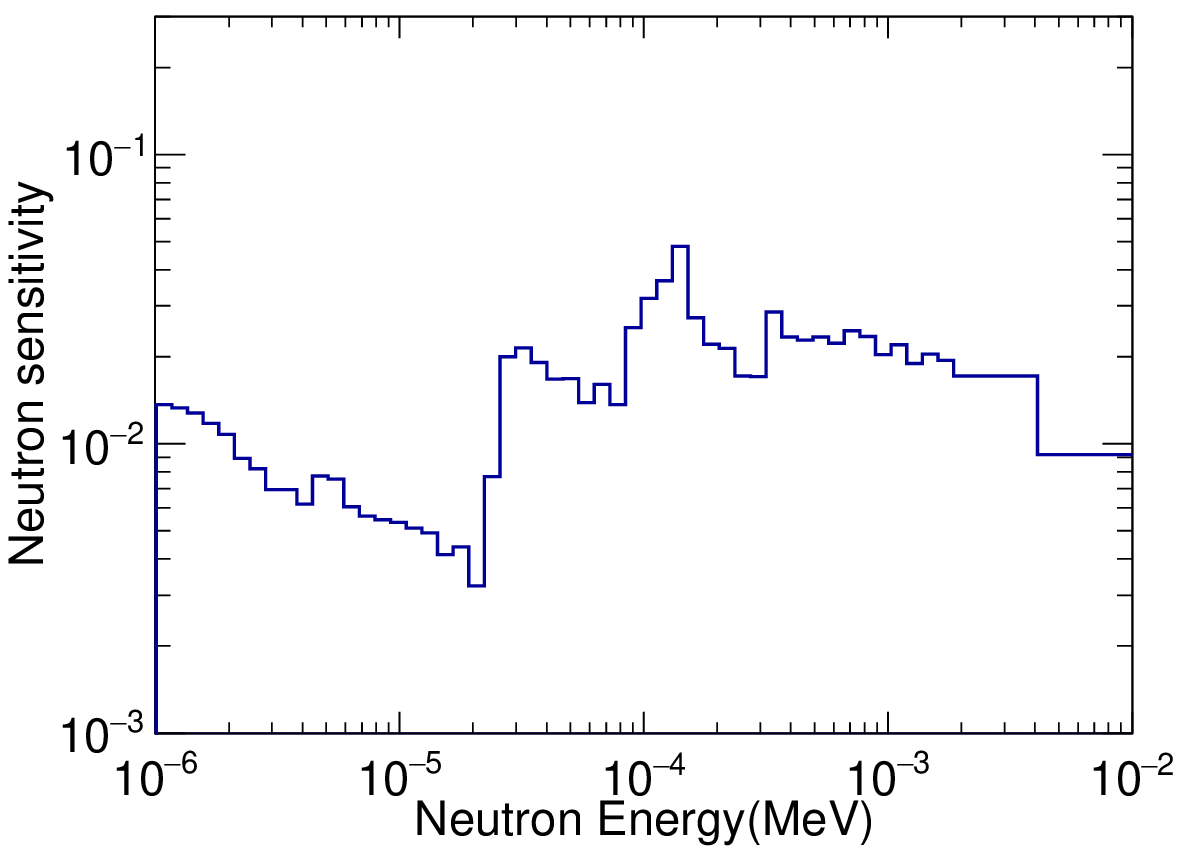}
\caption{(Color online) Neutron sensitivity of the GTAF}
\label{fig:Neutron sensitivity}
\end{figure}

The impact of scattered neutrons on the neutron capture measurement can be estimated by means of Monte Carlo simulations or determined through the use of pure scatterers. To quantify neutron sensitivity of the GTAF array, we used the carbon sample in which the dominant neutron process is elastic scattering, yielding negligible capture $\gamma$ rays.
The experimental (n,n) reaction yields and the theoretical yields are derived from Eq.(\ref{eqnn1},\ref{eqnn}), 
\begin{small}
\begin{eqnarray}
Y^{exp}_{n,n}(E_n)=\frac{C(E_n)-B(E_n)}{N \Phi (E_n)}
\label{eqnn1}
\end{eqnarray}
\end{small}
and 
\begin{small}
\begin{eqnarray}
Y^{th}_{n,n}(E_n)=(1-e^{N \sigma_t(E_n)}) \frac{\sigma_n(E_n)}{\sigma_t (E_n)}
\label{eqnn}
\end{eqnarray}
\end{small}

where the $C(E_n)$ and $B(E_n)$ represent the total counting rate and background of the carbon sample, respectively. The scattering neutron cross section ($\sigma_n(E_n)$) and the total neutron cross section ($\sigma_t(E_n)$) from the evaluated cross sections in the ENDF/B-VIII.0 library~\cite{bib:ENDF8}. The neutron sensitivity is then determined by comparing these experimental yields with the theoretical predictions. For the analysis conditions of $3.5 \text{MeV} \textless E_{sum} \textless 7 \text{MeV}$ and $m_{cr} \textgreater 2$, the neutron sensitivity is shown in Fig.\ref{fig:Neutron sensitivity}. It ranges from about 1.5\% at low energies to 3\% above 100 eV. The significant peaks at 30 eV and 100 eV correspond to resonances in the Ba and F capture cross sections within the BaF$_2$ detectors, temporarily increasing the sensitivity. Overall, the effect of neutron sensitivity on the $^{165}$Ho(n, $\gamma$) cross-section measurements is small, with an uncertainty of less than 0.5\% under these conditions.

\vspace{-1.0cm}

\subsection{Uncertainty}
The overall uncertainties considered in this work are summarized in Table \ref{tab:uncertainty}.
The $^{165}$Ho samples exhibited a uniform mass distribution with an uncertainty of 1\%. The $^{nat}$Dy impurity introduced a 0.5\% uncertainty in the $^{165}$Ho isotopic abundance. The uncertainty of the normalization factor based on the 12.7 eV saturated resonance peak is less than 2\%. The neutron flux shape simulated efficiency is 3\% uncertainty from 0.3 eV to 150 keV~\cite{2025chen}. Neutron sensitivity, assessed at $\textless$0.5\%, implied minimal scattered neutron background, with an estimated uncertainty of $\textless$1\%, as no significant scattering-to-capture ratio effects were observed. Detection efficiency and dead time losses contributed uncertainties of 2.5\% and 2\%, respectively~\cite{2021Zhang}.
The sum of these components yields an overall systematic uncertainty of 5.1\% for GTAF results.
\begin{table}[!htb]
\caption{Summary of relative uncertainty, sample uncertainty, normalization factor in this work.}
\label{tab:uncertainty}
\begin{ruledtabular}
\begin{tabular*}{8cm} {@{\extracolsep{\fill} } cc}
Source	& Uncertainty(\%) \\
\hline
mass distribution	             & 1         \\
Isotopic abundance	             & 0.5       \\
normalization factor	         & 2         \\
neutron energy    	             & 0.8(\textless1keV)\\
scattered neutrons	             & 1         \\
neutron flux                 	 & 3         \\
detection efficiency     	     & 2.5       \\
dead time losses	             & 2         \\
\hline
Total     	                     & 5.1       \\
\end{tabular*}
\end{ruledtabular}
\end{table}

\vspace{-0.5cm}
\section{RESULTS AND DISCUSSION}
\subsection{Capture yield}
The neutron capture yield of $^{165}$Ho was analyzed using the SAMMY code~\cite{OakRidgeReport} with the Reich-Moore approximation in the resolved resonance region (RRR). The initial resonance parameters (E$_R$, $\Gamma_{\gamma}$, $\Gamma_n$, scattering radius $R^l$, and spin groups $J$) were taken from the ENDF/B-VIII.0 data library with a 10\% input uncertainty~\cite{2011Chadwick}. The capture yield was fitted using a Bayesian approach incorporating covariance matrix, allowing adjustments to resonance energy and resonance width~\cite{2025wang}. The Back-n resolution function, incorporating multiple scattering corrections, flight path errors, neutron energy broadening, and Doppler effects, was integrated into SAMMY based on the calculations of Jiang {\it et al.}~\cite{2021Jiang} resonance parameters were refined through iterative fitting.

\begin{figure}[!htb]
\includegraphics[width=1.0\linewidth]{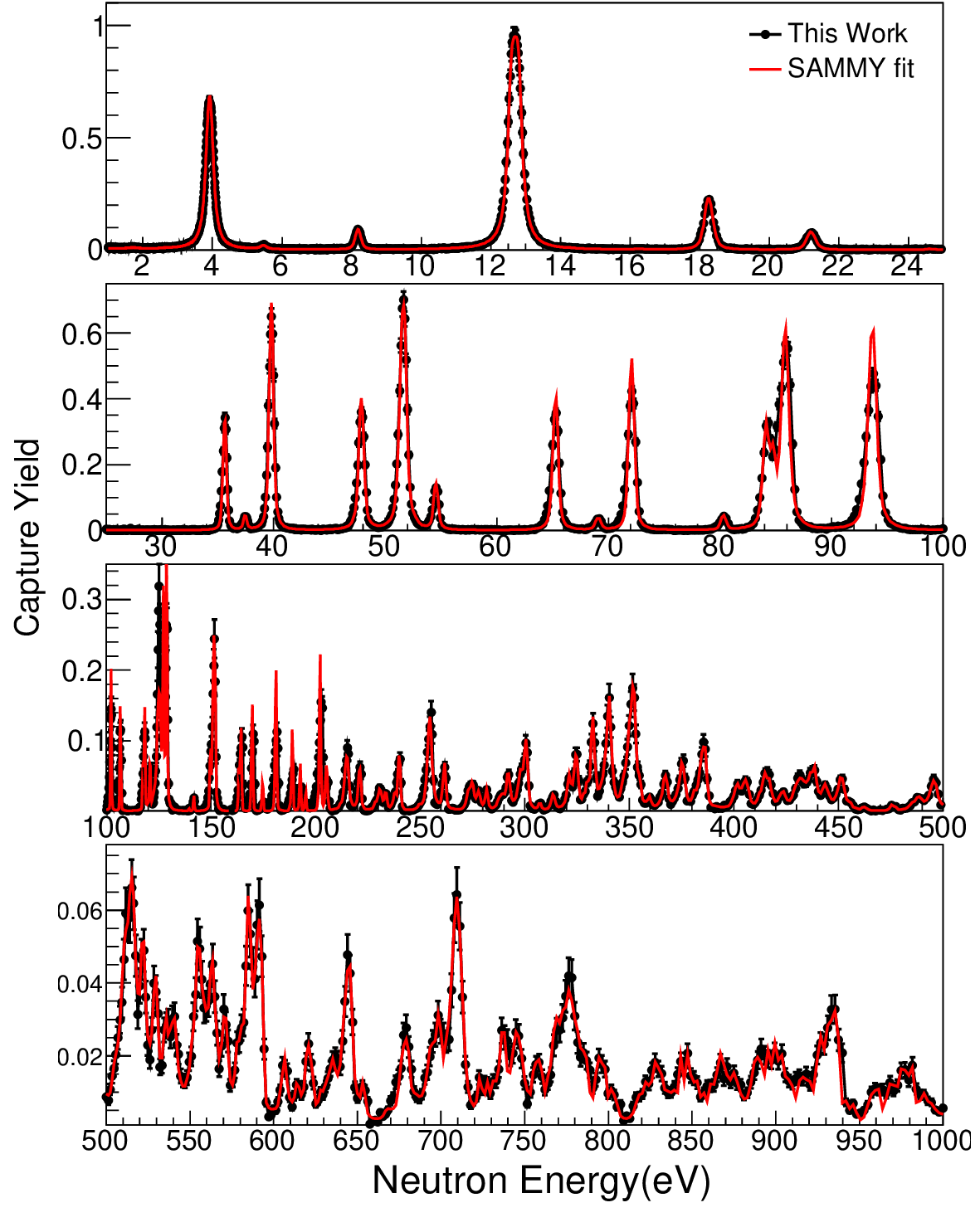}
\vspace{-0.4cm}
\caption{(Color online) The resonance fitting results for $^{165}$Ho capture yields below 1.0 keV. The black solid data points correspond to the experimental capture yield and the red solid line corresponds to the performed by SAMMY fit. The data align well with ENDF/B-VIII.0 below 100~eV, with deviations above this energy due to resolution limitations.}
\vspace{-0.5cm}
\label{fig:SAMMYfit}
\end{figure}

The analysis included contributions from $^{165}$Ho and seven isotopes of $^{nat}$Dy impurities. Despite the low $^{nat}$Dy concentration, isotopes such as $^{163}$Dy (1.7 eV, 16.2 eV) and $^{162}$Dy (5.43 eV) exhibited detectable high-cross-section resonances. 
The $^{nat}$Dy yields weighted by isotopic abundance and normalized to the 12.7 eV $^{165}$Ho saturation peak using Eq.(\ref{eq3}). 
The 12.7 eV resonance was selected for normalization as it is a well-resolved, saturated resonance in $^{165}$Ho, providing a robust reference for absolute calibration of capture yields.
The results of $^{165}$Ho capture yields and SAMMY fit below 1.0 keV are shown in Fig.\ref{fig:SAMMYfit}. 
Above 100 eV, the fitting results are not satisfactory due to the energy resolution of the back-n beam and the effect of impurities. 
However, all resonance peaks were well resolved in the range from 1 eV to 100 eV with energies consistent from the evaluated data of ENDF/B-VIII.0 library.

\vspace{-0.2cm}
%\vspace{-0.5cm}

\subsection{Resonance kernel}
\begin{figure}[!htb]
\includegraphics[width=.9\linewidth]{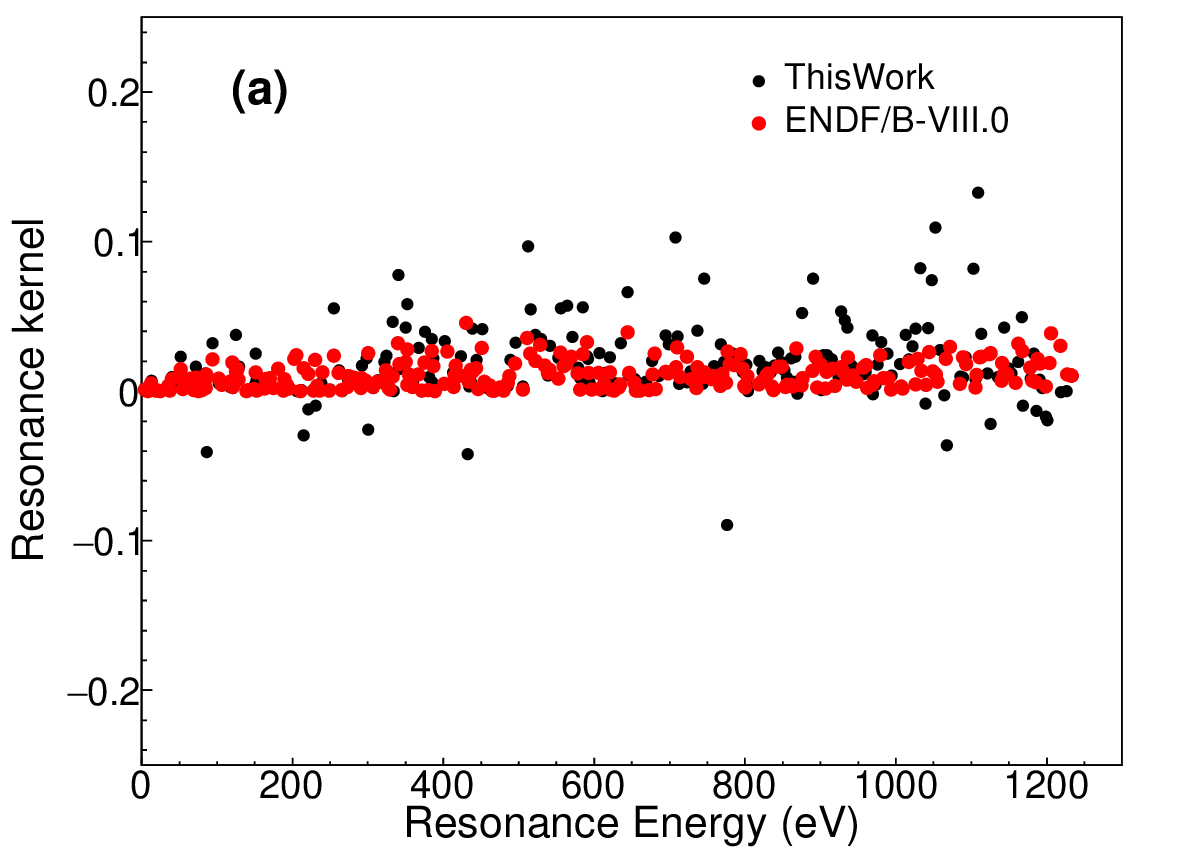}
\includegraphics[width=.9\linewidth]{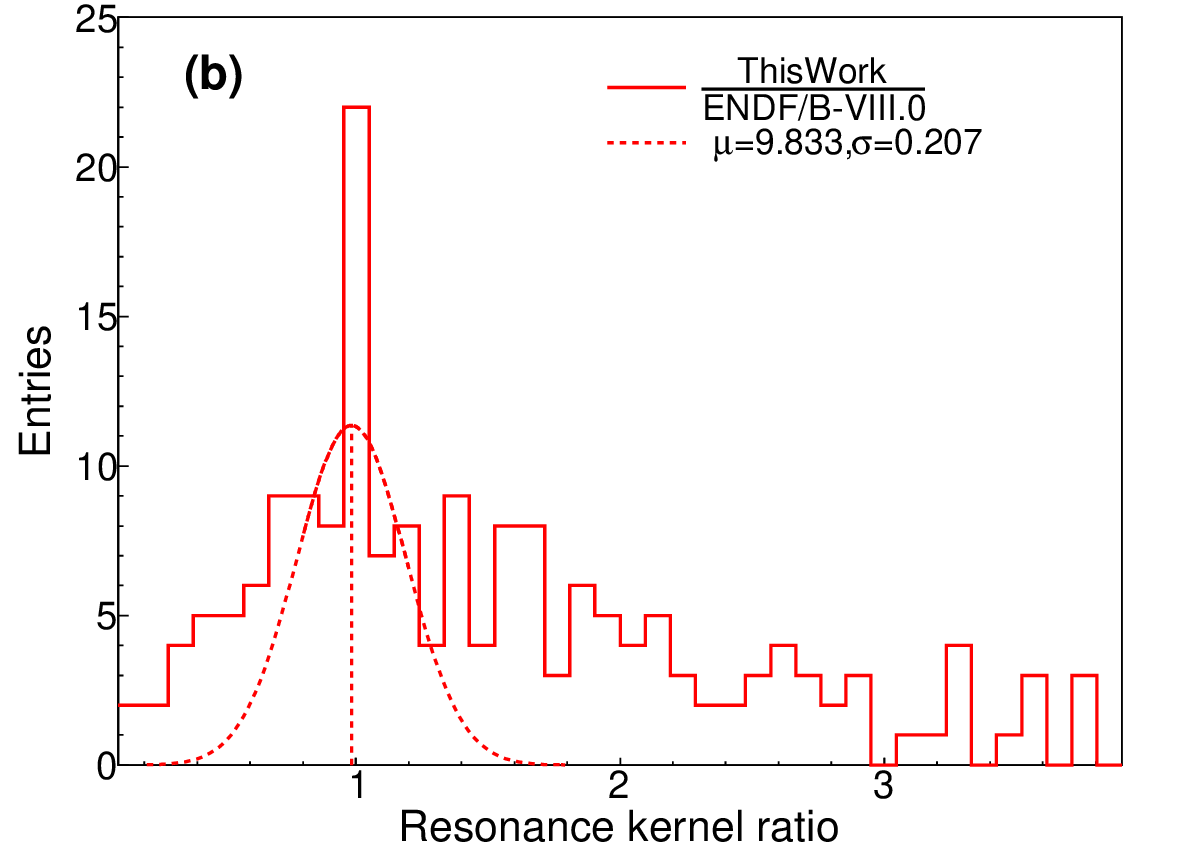}
\caption{(Color online) (a) Resonance kernels determined in the present work (black solid circles) plotted against resonance energy and compared with the evaluated data set ENDF/B-VIII.0 (red solid circles).
(b) Histogram of the resonance-kernel ratios (present work divided by ENDF/B-VIII.0); the red dashed curve represents a Gaussian fit to the distribution (fit parameters $\mu$ and $\sigma$ are quoted in the legend).}
\label{fig:ResonanceKernel}
\end{figure}

The resonance kernel $R_k$, which correlates the neutron width $\Gamma_n$ with the gamma width $\Gamma_\gamma$, is a key indicator for validating measurement accuracy in neutron capture experiments. It is defined as:
%\vspace{-0.3cm}
\begin{small}
\begin{eqnarray}
R_k=g \Gamma_n \Gamma_{\gamma}/(\Gamma_n +\Gamma_{\gamma})
\label{eq4}
\end{eqnarray}
\end{small}
where g is the spin statistical factor, given by $g=(2J+1)/((2s+1)(2I+1))$, with $J$ is the resonance spin, $I$ is the target nucleus spin, and $s = 1/2$ as the neutron spin. 
The resonance kernels of the GTAF and ENDF/B-VIII.0 were extracted up to 1.0 keV, and compared in Fig.\ref{fig:ResonanceKernel}. The data were analyzed with a Gaussian distribution, where the mean ($\mu$) reflects accuracy relative to ENDF/B-VIII.0, and the variance ($\sigma$) indicates the spread of the data.

As shown in Fig.\ref{fig:ResonanceKernel}(a), some experimental resonance kernels (e.g., at 24.7 eV and 75.8 eV) are negative, despite their presence in ENDF/B-VIII.0.
These resonances are not distinctly observed in the experimental data, leading to the SAMMY code fitting them with negative $\Gamma_\gamma$ widths. 
This could be related to the thickness of the $^{165}$\text{Ho} target in the experiment. For such a thin target, the neutron absorption probability is limited, especially for resonances with lower cross sections or narrower widths. This may result in insufficient signal-to-noise ratios for some resonances, making them indistinguishable from the background or overlapping with adjacent resonance features, thus failing to be clearly resolved in the experimental data.
Additionally, some resonance kernels exhibit significant deviations, largely resulting from the resonance overlap due to the limitations of the experimental equipment and energy resolution of the beam in separating closely positioned resonances. 

In general, the experimental results are in good agreement with those from the ENDF/B-VIII.0 library, especially for energies below 100 eV. Thus, only the resonance parameters up to 100 eV are reported in this work. The obtained values of the resonance parameters are provided in Table \ref{tab:Resonance parameter}.

\vspace{-0.3cm}

\vspace{-0.3cm}
\begin{table*}[!htb]
\caption{Resonance parameter results for This work, ENDF/B-VIII.0 and Asghar{\it et al.} (1968) result.}
\label{tab:Resonance parameter}
\begin{ruledtabular}
\begin{tabular*}{15cm} {@{\extracolsep{\fill} } c|cccc|cccc|cccc}
\multirow{2}*{J$^{\pi}$} & \multicolumn{4}{c|}{ENDF/B-VIII.0} & \multicolumn{4}{c|}{This Work} & \multicolumn{4}{c}{Asghar {\it et al.}(1968)}\\
    \cline{2-13}
    &$E_{\text{R}}$ (eV) &$\Gamma_{\gamma}$(meV) &$\Gamma_n$(meV) &$k$&$E_{\text{R}}$ (eV) &$\Gamma_{\gamma}$(meV) &$\Gamma_n$(meV)&$k$&$E_{\text{R}}$ (eV) &$\Gamma_{\gamma}$(meV) &$\Gamma_n$(meV)&$k$\\
\hline
4.0  &  3.914	& 85.70	  &  2.132	  & 1.17   & 3.913$\pm$0.001& 92.59$\pm$0.28    & 2.132$\pm$0.01  &1.172$\pm$0.005     &3.92&  -$^{\footnotemark[1]}$   &  -    		 & -       \\
3.0  &  8.174	& 90.30	  &  0.187	  &0.082   & 8.187$\pm$0.025&  91.46$\pm$0.35   & 0.204$\pm$0.02  & 0.089$\pm$0.008   &	8.14	&  -		 	 &  0.17$\pm$0.06    & -       \\
4.0  &  12.69	& 84.00	  &  10.31	  &5.17    & 12.68$\pm$0.02 &  88.32$\pm$0.21   & 11.45$\pm$2.30  & 5.70$\pm$1.01      &	12.63	&  -			 &  15.0$\pm$0.15    & -       \\
3.0  &  18.25	& 78.10	  &  0.95	  &0.41    & 18.24$\pm$0.02 &  89.27$\pm$0.53   & 1.58$\pm$3.50   & 0.68$\pm$1.48      &	18.08	&  -		     &  1.35$\pm$0.14    &-        \\
4.0  &  21.19	& 68.00	  &  0.52	  &0.29    & 21.17$\pm$0.04 &  89.63$\pm$0.84   & 0.60$\pm$0.04   &0.337$\pm$0.02      &	20.97	&  -		     &  0.63$\pm$0.2     &         \\
3.0  &  24.79	& 84.00	  &  0.02	  &0.001   & -				&  -  				& -   			  &-				   &	-		&  -		     &  -   			 & -       \\
3.0  &  35.33	& 73.60	  &  8.69	  &3.40    & 35.57$\pm$0.34 &  85.59$\pm$3.52   & 7.01$\pm$0.72   & 2.83$\pm$0.27      &	35.42	&  72.6$\pm$5.5  & 7.4$\pm$0.1   & 2.94$\pm$0.04   \\
4.0  &  37.36	& 83.00	  &  0.50	  &0.28    & 37.41$\pm$0.07 &  87.21$\pm$0.68   & 0.54$\pm$0.08   & 0.30$\pm$0.04      &	-		&  -		 	 &  -    	     & -        	    \\
4.0  &  39.67	& 88.00	  &  16.80	  &7.93	   & 39.76$\pm$0.12 &  95.28$\pm$0.67   & 18.14$\pm$0.24  & 8.57$\pm$0.09      &	39.59	&  90.1$\pm$4.5  & 19.5$\pm$0.2  & 9.02$\pm$0.11    \\
3.0  &  47.80	& 92.00	  &  28.22	  &9.45    & 47.81$\pm$0.04 &  94.54$\pm$0.98   & 16.74$\pm$2.25  & 6.22$\pm$0.71	   &	47.63	&  90$\pm$2.1	 &  26.9$\pm$0.3 &6.22$\pm$0.10	    \\
3.0  &  51.55	& 85.00	  &  56.57	  &14.86   & 51.60$\pm$0.07 &  89.61$\pm$0.83   & 82.75$\pm$5.80  & 18.82$\pm$0.69     &	51.36	&  79.0$\pm$2.6  & 54.0$\pm$1.2  & 14.03$\pm$0.26   \\
4.0  &  54.42	& 84.00	  &  2.40	  &1.31    & 54.49$\pm$0.12 &  87.23$\pm$2.56   & 3.73$\pm$0.35   & 2.013$\pm$0.18	   &	54.22	&  -			 &  3.3$\pm$0.8  & - 		  		\\
4.0  &  65.15	& 77.00	  &  18.66	  &8.45    & 65.23$\pm$0.23 &  86.54$\pm$3.52   & 17.04$\pm$1.92  & 8.01$\pm$0.75      &	64.96	&  74.6$\pm$2.0  & 20.6$\pm$0.3  & 9.08$\pm$0.12   	\\
4.0  &  68.91	& 89.00	  &  1.10	  &0.61    & 69.01$\pm$0.12 &  87.47$\pm$0.79   & 1.17$\pm$0.06   & 0.64$\pm$0.03	   &	68.75	&  -		 	 &  1.0$\pm$0.3  & -   		  		\\
4.0  &  71.93	& 74.00	  &  20.44	  &9.01    & 72.10$\pm$0.34 &  86.70$\pm$9.20   & 18.09$\pm$3.04  & 6.54$\pm$1.16      &	71.73	&  74.9$\pm$1.8  & 23.7$\pm$0.3  & 10.12$\pm$0.11   \\
3.0  &  75.08	& 84.00	  &  0.095	  &0.04    & -				&	-    		    & -	 			  &  -   			   &   -		&  -			 &  -   		 &  -          		\\
4.0  &  80.10	& 82.00	  &  1.55	  &0.85    & 80.13$\pm$0.06 &  80.29$\pm$2.10	& 7.11$\pm$0.46   & 3.67$\pm$0.21 	   &79.92	    & -				 &  1.6$\pm$0.7  & -  		  		 \\
4.0  &  83.80	& 67.00	  &  13.51	  &6.32    & 84.14$\pm$0.26 &  84.11$\pm$7.34	& 21.37$\pm$2.80   & 9.59$\pm$1.02     &83.76	    &  66.7$\pm$4.3  & 11.2$\pm$0.2  & 5.39$\pm$0.10  	 \\
3.0  &  84.73	& 84.00	  &  5.60	  &2.29    & - 				&  - 				& - 			  &	-				   &  -	        &  -             &               &       	  		 \\
3.0  &  85.80	& 84.00	  &  37.37	  &11.31   & 86.03$\pm$0.32 &	83.83$\pm$0.45  & 45.46$\pm$3.56 & 12.89$\pm$0.65 	   & 85.5	    &  *$^{\footnotemark[2]}$&  *    &   * 		   		\\
4.0  &  93.63	& 79.00	  &  73.77	  &21.45   & 93.78$\pm$0.22 &   93.78 $\pm$0.76 & 45.23$\pm$6.78 & 16.67$\pm$1.63 	   &93.3	    &  *			 &  *  			 &  *         		\\
\end{tabular*}
\end{ruledtabular}
\footnotetext[1]{Data not measured on the experiments.}
\footnotetext[2]{The results were measured in a capture experiment, but could not be displayed due to printing.}
\end{table*}
\nocite{*}

\subsection{Statistical analysis of resonances}
\subsubsection{Average Level Spacing}
\begin{figure}[!htb]
\includegraphics
  [width=0.8\hsize]
  %{Fig1_TbCaptureYield.eps}
  {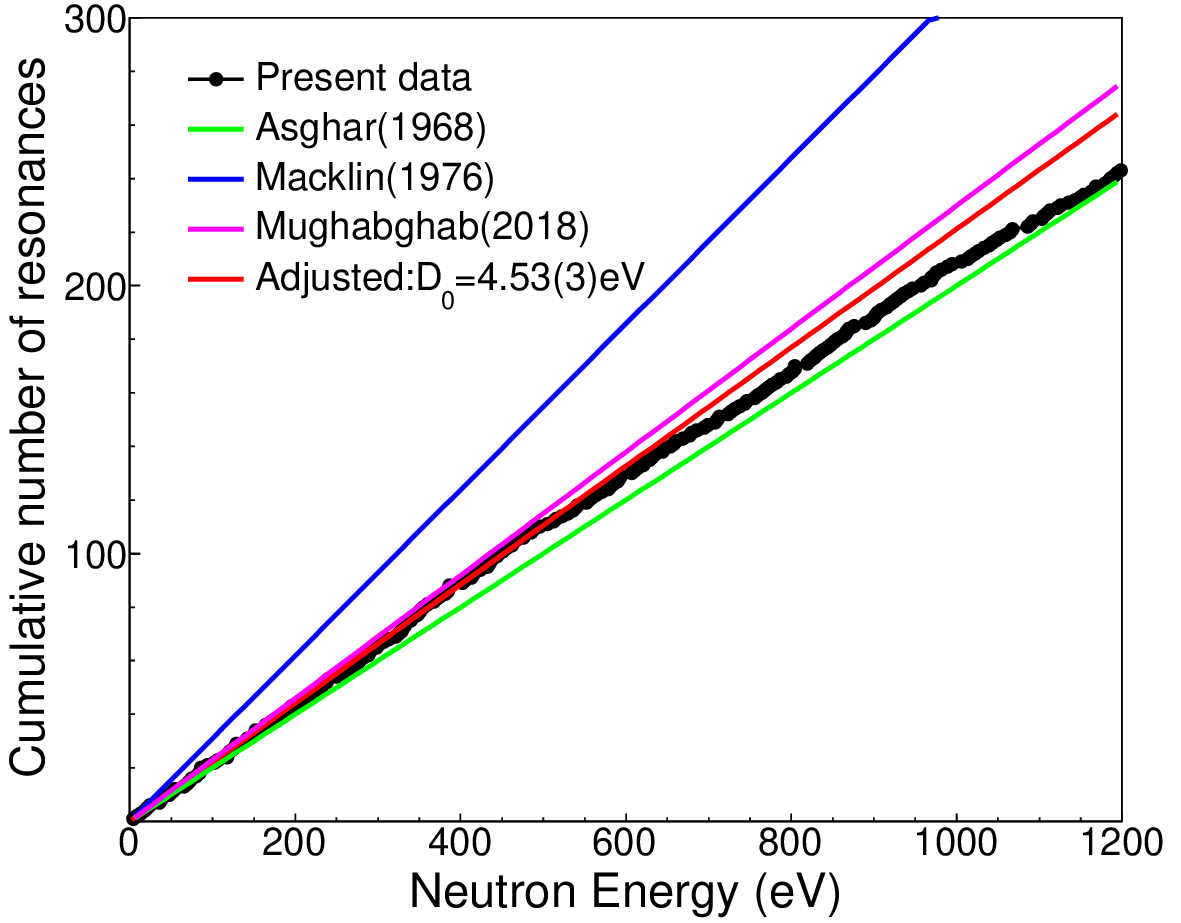}
  \caption{Cumulative number of levels as a function of neutron energy.}
  \label{fig:Cumulative}
\end{figure}

The average level spacing, defined as the mean energy difference between consecutive resonance levels, is a fundamental quantity that reflects the density of nuclear states at the neutron separation energy. 
The cumulative plot of experimentally observed resonance counts as a function of neutron energy is depicted in Fig. \ref{fig:Cumulative}.
Analysis of the cumulative resonance indicates discrepancies exceeding \SI{100}{\electronvolt } from the reported data in the Ref.~\cite{bib:Mughabghab}. Moreover, the proportion of missing resonances steadily increases with the neutron energy increases, a phenomenon potentially attributable to statistical fluctuations.
As shown in Fig. \ref{fig:Cumulative}, a linear fit for the average level spacing below \SI{100}{\electronvolt} a value of $D_0$ =\SI{4.53(3)}{\electronvolt}. 

\begin{figure}[!htb]
\includegraphics
  [width=0.85\hsize]
  %{Fig1_TbCaptureYield.eps}
  {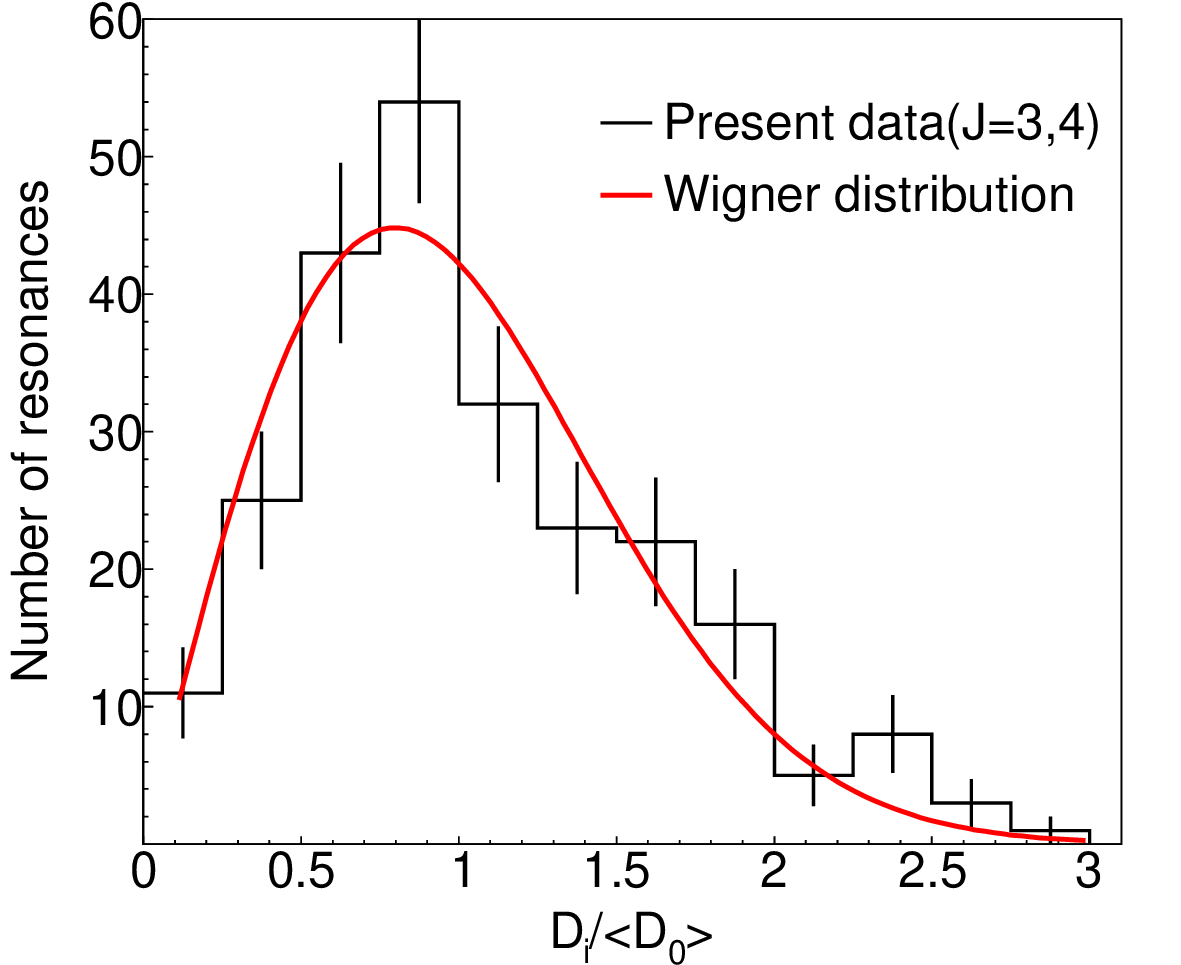 }
  \caption{Experimental and theoretical distribution of the next neighbouring resonance distances)}
\label{fig:WignerDistribution}
\end{figure}

In analyzing level spacing for $^{165}$Ho resonances, we evaluated the consistency with Wigner-Dyson statistics~\cite{bib:Wigner}, which  provides a mathematical prediction of the level spacing distribution in chaotic systems, accurately reproducing experimental observations. The probability distribution of these level spacing is expected to follow a distribution as Eq.(\ref{eq:3}).
\begin{equation}\label{eq:3}
p(x)dx=\frac{\pi x}{2} exp(-\frac{\pi x^2}{4})dx,\quad with \quad x=\frac{D_i}{\langle D \rangle}
\end{equation}
where $D_i$ is the energy level spacing of a series of N consecutive resonances with energy $E_i$, expressed as $D_i=E_i-E_{i-1}$.
This value is compared to the expected theoretical Wigner distribution, as illustrated in Fig. \ref{fig:WignerDistribution}, serving as a test for the consistency of the observed spin experimental level spacing distribution.

\vspace{-0.2 cm}
\vspace{-0.2 cm}

\vspace{-0.2 cm}
%\vspace{-2.5 cm}

\subsubsection{Neutron Width}
The neutron width $\Gamma_n$, representing the partial width associated with neutron emission in the decay of a compound nuclear resonance, is a key parameter in understanding the dynamics of neutron capture reactions. In this study, $\Gamma_n$ values were extracted for 18 resolved resonances in $^{165}\text{Ho}$ in the neutron energy range of 1 eV to 100 eV. 
The distribution of reduced neutron widths, defined as  $\Gamma_n^0$=$\Gamma_n$ / $\sqrt{E_n}$, where $E_n$ is the resonance energy, was analyzed to probe the statistical properties of the neutron channel. 

To characterize the statistical distribution of the reduced neutron widths, the normalized widths x = $\Gamma_n^0$ / $\langle \Gamma_n^0 \rangle$ were compared to the Porter-Thomas $(P-T)$ distribution~\cite{bib:Porter1956}, a chi-squared distribution with one degree of freedom, given by:
\begin{equation}\label{eq:8}
P_{PT}(x)dx=\frac{1}{\sqrt{2\pi c}} e^{-\frac{x}{2}dx}.
\end{equation}

\begin{figure}[!htb]
\includegraphics
  [width=0.95\hsize]
  %{Fig1_TbCaptureYield.eps}
  {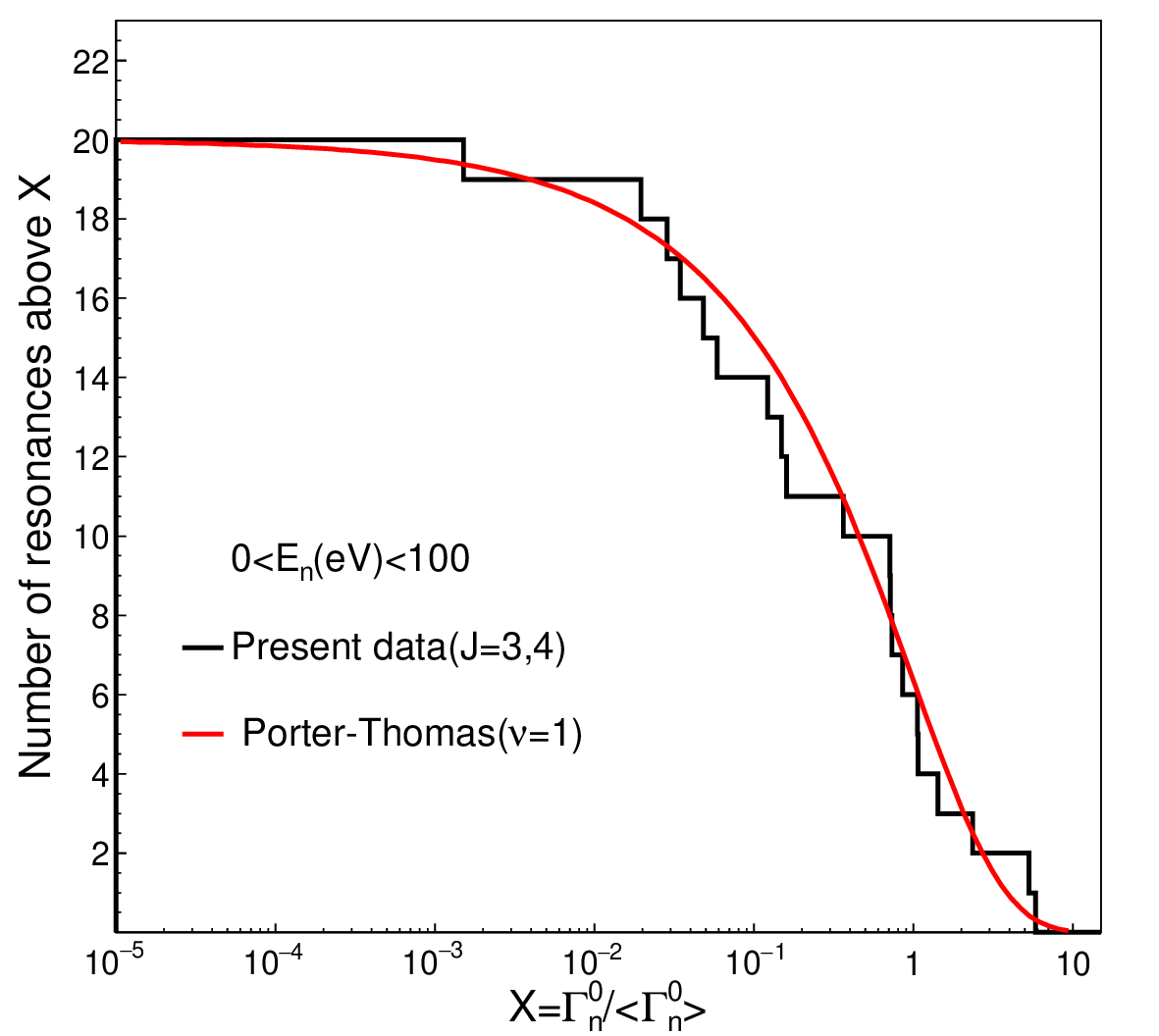}
  \caption{Comparison between the experimental and expected integral distributions of the neutron widths}
\label{fig:DistributionsOfNeutronWidths}
\end{figure}

Due to the lack of resonances above \SI{100}{\electronvolt}, the distribution of neutron widths is exclusively examined within the range of \SI{1}{\electronvolt} to \SI{100}{\electronvolt}. Figure~\ref{fig:DistributionsOfNeutronWidths} illustrates the experimentally observed and computationally derived resonance numbers, with $\Gamma_n^0 > x_i \langle \Gamma_n^0\rangle$ as a function of $x_i$. Upon observation, it is apparent that the experimental data shows remarkable alignment with the expected behavior for $J=1$ resonances. This result serves to substantiate the consistency of resonance parameters within the energy region characterized by minimal resonance overlap.

\vspace{-0.5cm}

\subsubsection{Average Radiative Width}
The average radiative width $\langle\Gamma_\gamma\rangle$ is a critical parameter for understanding the electromagnetic deexcitation processes in neutron induced reactions, representing the mean partial width for radiative decay of the compound nuclear states formed in neutron capture. 
In this study, $\Gamma_\gamma$ values were extracted for 18 resolved s wave resonances in the neutron energy range of 1 eV to 100 eV. 
The average radiative width was calculated as $\langle \Gamma_\gamma \rangle$ = 88.10 $\pm$ 1.98 $\text{meV}$, based on the arithmetic mean of the individual $ \Gamma_\gamma$ values across the resonance data. 

\begin{figure}[!htb]
\includegraphics
  [width=0.95\hsize]
  %{Fig1_TbCaptureYield.eps}
  {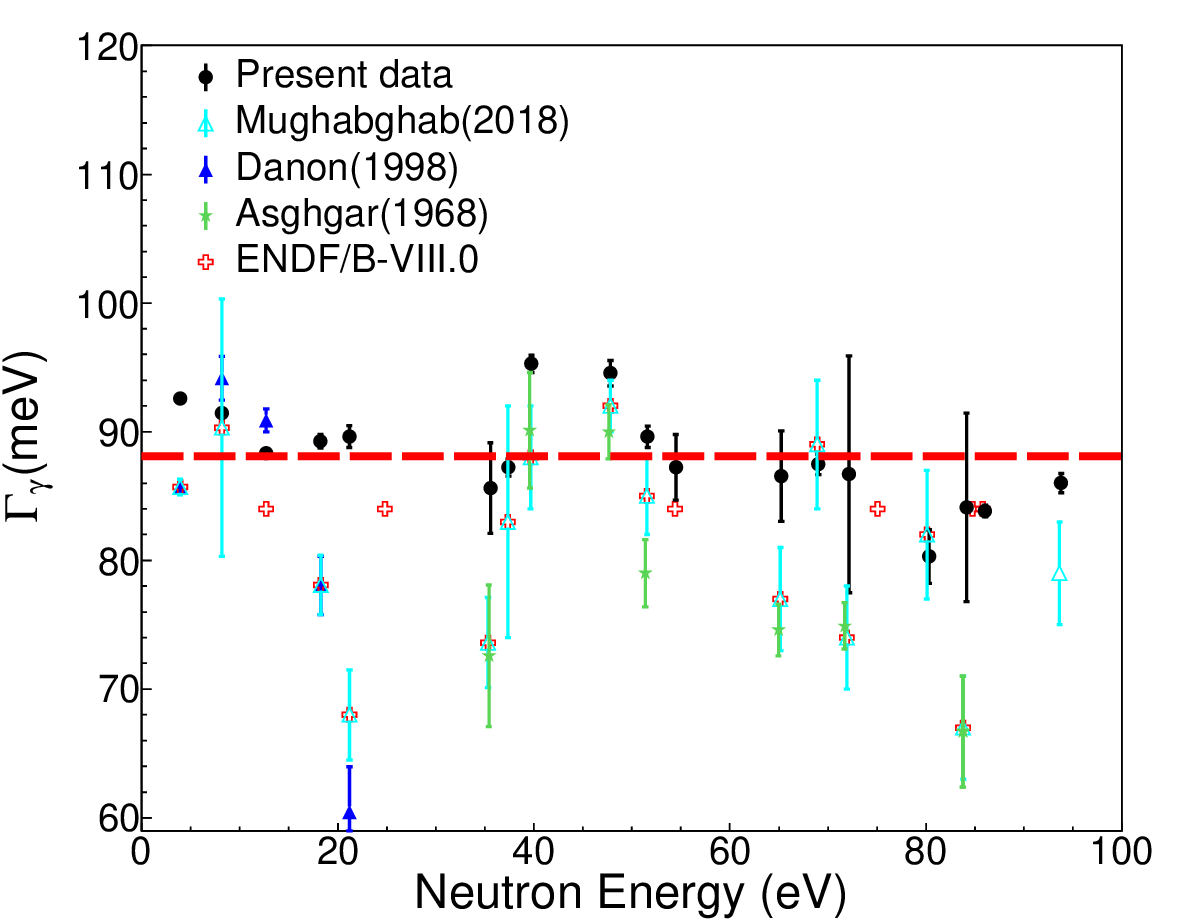}
  \caption{Fitted values for the radiative widths below \SI{100}{\electronvolt}. The red-dashed line corresponds to the arithmetic mean value $\langle \Gamma_{\gamma} \rangle =$ \SI{88.1 \pm 1.98}{m\electronvolt}.}
\label{fig:RadiativeWidth}
\end{figure}

Figure~\ref{fig:RadiativeWidth} presents the fitted values for the radiative widths below 100 eV, with the red-dashed line corresponding to the arithmetic mean value. This figure also includes comparisons with data from other sources such as Mughabghab(2018)~\cite{bib:Mughabghab}, Danon {\it et al.} (1998)~\cite{1998Danon}, Asghar {\it et al.} (1968)~\cite{1968Asghar}, and the ENDF/B-VIII.0 library~\cite{bib:ENDF8}, providing a context for evaluating the consistency of the measured average radiative width with previous studies.

\vspace{-0.5cm}

\subsubsection{Neutron Strength Function}
The neutron strength function $S_0$ is a fundamental quantity in neutron capture reactions, reflecting the strength of the neutron interaction with the target nucleus. The definition of the S-wave resonance (l=0) is the average reduced neutron width per unit energy interval divided by the average level spacing,  $S_0$ = $\langle g\Gamma_n^0 \rangle$ / $\langle D_0 \rangle$ . 
This implies that the $S_0$ can be calculated based on the slope of the cumulative distribution of $g\Gamma_n^0$, which acts as a neutron energy function. 

\begin{figure}[!htb]
\includegraphics
  [width=0.95\hsize]
  %{Fig1_TbCaptureYield.eps}
  {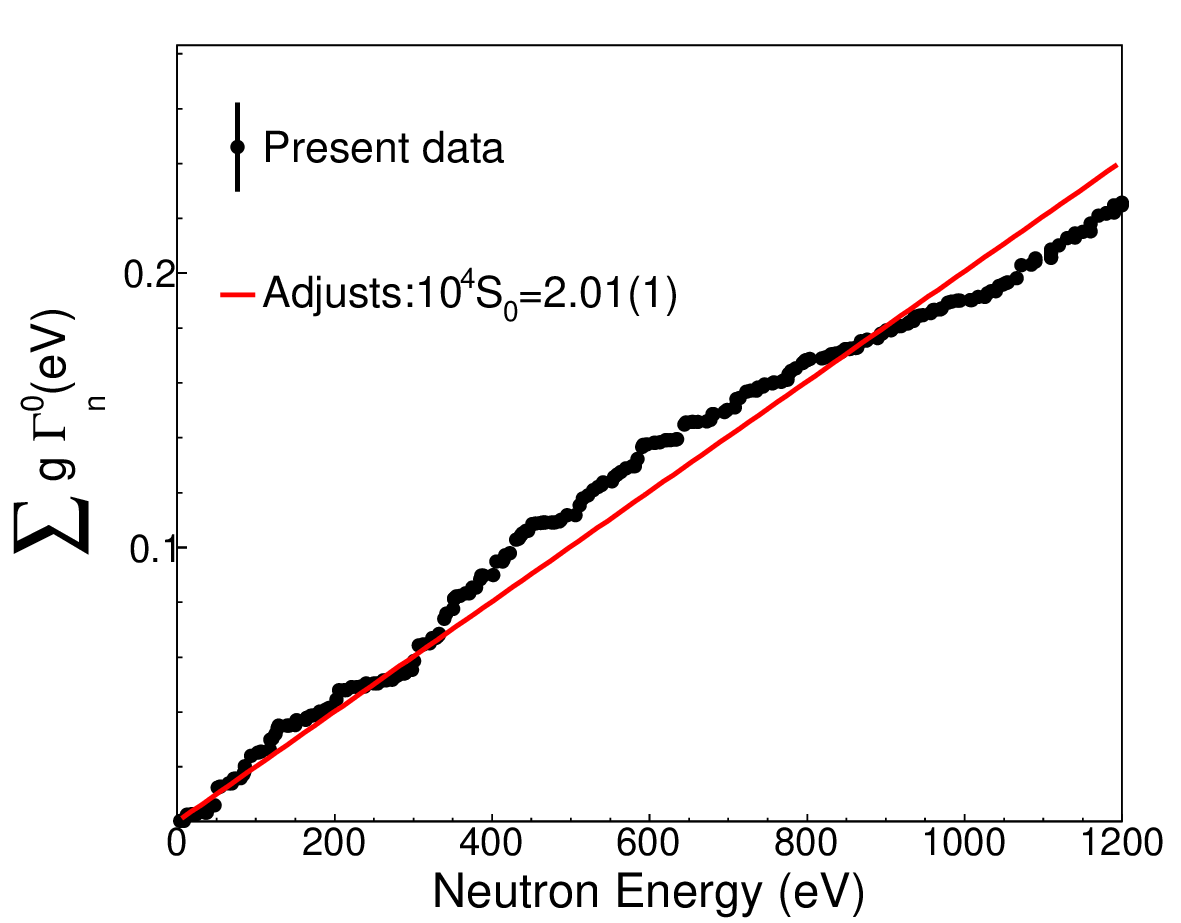}
  \caption{Cumulative sum of the reduced neutron widths as a function of neutron energy.}
\label{fig:NeutronStrengthFunction}
\end{figure}

The distribution, depicted in Fig.~\ref{fig:NeutronStrengthFunction} in black point, corresponds to a linear fit, as indicated by the red line, $10^{-4}S_0=\SI{2.01 \pm 0.01}{}$.
The neutron strength function was calculated across the full energy range. Despite the absence of resolved resonances above \SI{100}{\electronvolt}, the calculations remain effective. This effectiveness arises because the strength function being an averaged quantity associated with the total resonance area. It is noteworthy that this quantity is independent of the number of resonances employed, even in cases involving resonance multiplets.
The average parameters resulting from the statistical analysis of the resonance parameters of the SAMMY fits are compared to those results reported by Muhabghab~\cite{bib:Mughabghab} and Marklin~\cite{1976Macklin} in table ~\ref{tab:AverageResonanceParameters}. 

\begin{table}[!htb]
\caption{Average resonance parameters of $^{165}$Ho(n,$\gamma$) determined in this work compared to the literatures' values.}
\label{tab:AverageResonanceParameters}
\begin{tabular*}{8cm} {@{\extracolsep{\fill} } lrrr}
\toprule
  & This Work  &Muhabghab  ~\cite{bib:Mughabghab}& Marklin ~\cite{1976Macklin} \\
\hline
$D_0(eV)$              & 4.53(3)  &4.35(15)    & 3.23(55) \\
$\Gamma_{\gamma}(meV)$ & 88.10(198)      &84.00(5)   & 76.00  \\
$10^{4}S_0$  & 2.01(1) &1.97(19)  & 1.33(14) \\
\hline\hline
\end{tabular*}
\end{table}

\section{summary and Conclusion}
The neutron capture measurement of $^{165}$Ho was performed at the CSNS Back-n facility using the GTAF detection system. 
The experiment covered neutron energies from 1 eV to 1.0 keV, with a particular focus on the resolved resonance region below 100 eV.  Capture yields were obtained under optimized experimental conditions, utilizing thresholds for total deposited energy $3.5 \text{MeV} \textless E_{sum} \textless 7\text{MeV}$ and crystal multiplicity ( $m_{cr} \textgreater 2$) to minimize background contributions from in-beam $\gamma$-rays, scattered neutrons, and detector-related reactions. Neutron sensitivity was evaluated to be less than 0.5\%, and the detection efficiency was estimated to be 14.3$\pm$0.06\%.

Resonance shape analysis was performed with the the R-matrix code SAMMY. Refined resonance parameters, such as resonance energy $E_R$, radiative width $\Gamma_\gamma$, and neutron width $\Gamma_n$ were obtained for incident neutron energies up to 100 eV. Comparison with the ENDF/B-VIII.0 library showed overall consistency, but a significant deviation was observed at 12.7 eV, where the measured capture cross section exceeded that at 3.9 eV—contrary to the trend in the evaluated data. This result aligns with earlier measurements by Asghar {\it et al.}, suggesting a need for reevaluation of the ENDF/B-VIII.0 data in this energy range.

Statistical analyses of the resonance parameters, the average level spacing below 100 eV was determined to be $D_0 = 4.53(3) \text{eV}$; the distribution of reduced neutron widths aligned with the Porter-Thomas distribution, validating the statistical behavior of the neutron channel; and the average radiative width for s-wave resonances was found to be $\langle \Gamma_\gamma \rangle$ = 88.10 $\pm$ 1.98 $\text{meV}$. Additionally, the neutron strength function $10^{-4}S_0=\SI{2.01 \pm 0.01}{}$ was derived, reflecting the strength of the neutron-nucleus interaction in $^{165}$\text{Ho}.

Future work could enhance data quality by incorporating $^{10}$\text{B} neutron absorbers into GTAF to reduce scattered neutron background and improve shielding against in-beam $\gamma$-rays; this upgrade would enable more precise measurements of higher-energy resonances (above 100 eV) that are currently limited by resolution. Overall, the present study provides updated nuclear data for $^{165}$Ho(n,$\gamma$) reaction that are essential for nuclear data evaluations, neutron capture application and statistical model validation.

%\vspace{-0.2cm}
\section{Acknowledgem}
\vspace{-0.1cm}
This work was supported by the National Natural Science Foundation of China (Grants Nos.12465024 and 12365018) and Natural Science Foundation of Inner Mongolia (Grants Nos. 2024ZD23, 2024FX30 and 2023MS01005) and the program of Innovative Research Team and Young Talents of Science and Technology in Universities of Inner Mongolia Autonomous Region (Grant No.NMGIRT2217 and NJYT23109),
and the Central Government Guidance for Local Science and Technology Development Funds Project(2025ZY0067).
We also thank the staff members of the Back-n white neutron facility (https://cstr.cn/31113.02.CSNS.Back-n) at the China Spallation Neutron Source (CSNS) (https://cstr.cn/31113.02.CSNS), for providing technical support and assistance in data collection and analysis.


\end{document}